\documentclass[prb,preprint]{revtex4-1} 

\usepackage{amsmath}  
\usepackage{amsfonts} 
\usepackage{graphicx} 

\usepackage{color,soul}

\begin{document}


\title{Young's Double-Slit Interference Demonstration with Single Photons}

\author{Bill J. Luo, Leia Francis, Valeria Rodr\'iguez-Fajardo,  and Enrique J. Galvez}
\email{egalvez@colgate.edu} 
\affiliation{Department of Physics and Astronomy, Colgate University, Hamilton, NY 13346, USA}

\author{Farbod Khoshnoud}
\affiliation{Electromechanical Engineering Technology Department, College of Engineering,
California State Polytechnic University, Pomona, CA 91768, USA}


\date{\today}

\begin{abstract}


The interference of single photons going through a double slit is a compelling demonstration of the wave and particle nature of light in the same experiment. Single photons produced by spontaneous parametric down-conversion can be used for this purpose. However, it is particularly challenging to implement due to coherency and resolution challenges.  In this article, we present a tabletop laboratory arrangement suitable for the undergraduate instruction laboratory that overcomes these challenges. The apparatus scans a single detector to produce a plot showing the interference patterns of single photons. We include experimental data obtained using this setup demonstrating double-slit and single-slit interference as well as quantum erasing through the use of sheet polarizers.

\end{abstract}

\maketitle 

\section{Introduction} 
In 1807 Thomas Young published a two-volume compendium of his lectures on many physics topics.\cite{Young} It included the description of a transformative experiment in physics: the interference of light passing through a double slit. Up to that point, Newton's hypothesis of light corpuscles was the prevailing understanding of light. 
Young's treatise on interference was important because it established the wave nature of light. 
It took a long time for Newton's hypothesis of corpuscles of light to be abandoned, but by the end of the 19th century, the wave nature of light was widely accepted. Then, in 1905, Albert Einstein proposed the concept of light made of quanta (later named \emph{photons}) in explaining the photoelectric effect.\cite{EinsteinPR05} 
It was initially thought to be a reversal back to Newton's theory of light,
but instead, it was the start of a new way of understanding light. 
Quantum mechanics provided the framework for this new (and contemporary) understanding, but not without the following conceptual challenge: Light interferes like a wave, but it is detected whole, like a particle. 
In 1927 Bohr proposed a way to deal with the apparent contradiction via the principle of complementarity,\cite{BohrN28} which, in a broad sense, states that a 
quantum object can exhibit either particle or wave behavior, but not both simultaneously.\cite{Folse} 

The purpose of this article is to demonstrate the phenomenon of complementarity in an undergraduate-accessible tabletop experiment involving the interference of single photons.
In Sec.~\ref{sec:theor} we present the theory that describes 
our experiment.
 We follow with a general description of the apparatus in Sec.~\ref{sec:app}. In Sec.~\ref{sec:results}, we show the results of several experiments: the double-slit case, the single-slit case, and the double-slit with polarization marker and its nulling via quantum erasing. We follow with Sec.~\ref{sec:cons}, which presents the experimental challenges that must be overcome to observe the phenomenon in such a simple setting. Conclusions follow in Sec.~\ref{sec:conc} and a list of parts is in the Appendix.

\section{Theory}\label{sec:theor}

\begin{figure}[h!]
\centering
\includegraphics[width=4in]{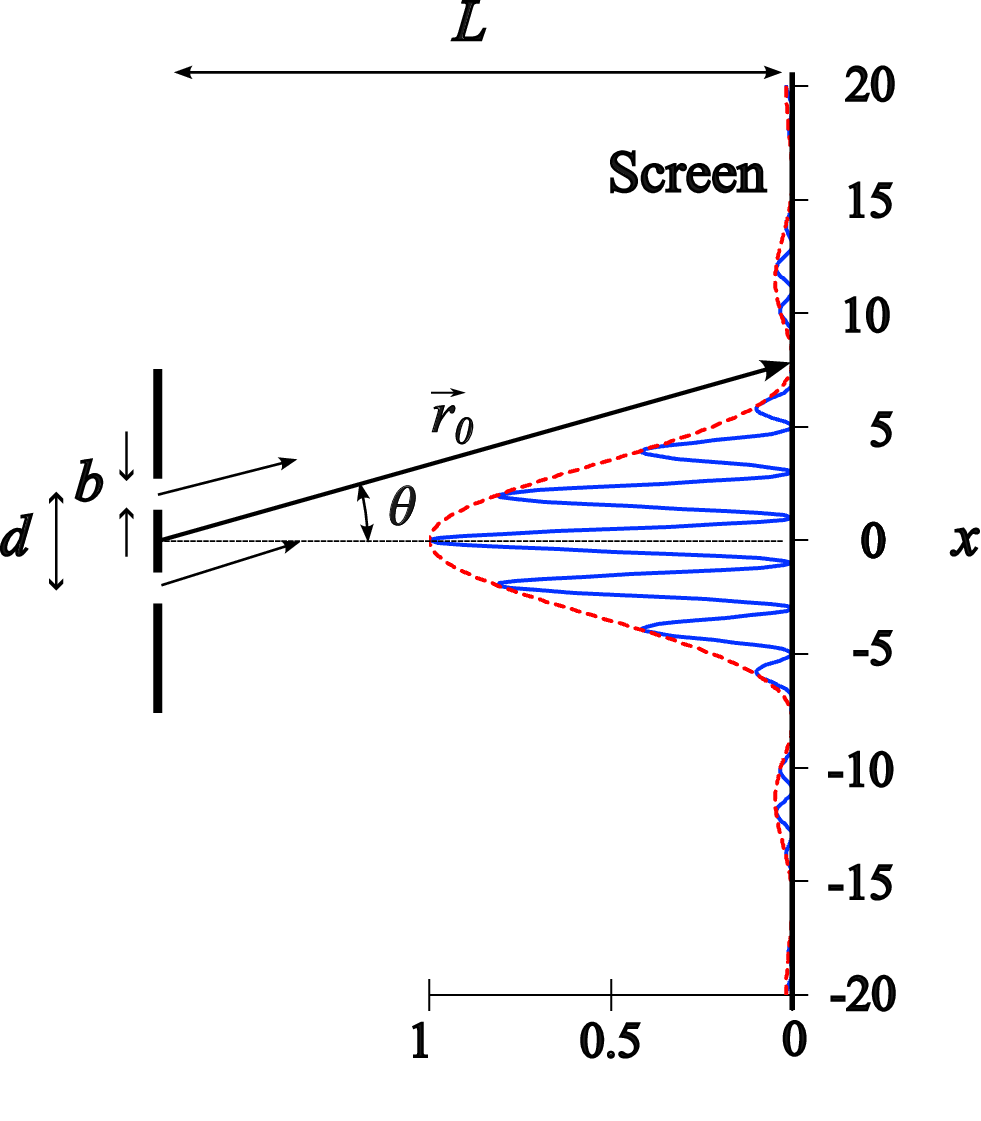}
\caption{(Color online) Schematic of a double-slit interference setup, and the interference pattern predicted by Eq.~(\ref{eq:IP}) (solid line), also showing the 
single-slit envelope (dashed line). For the case shown (not to scale), the distance between maxima is $\delta x=\lambda L/d=2$ units on the screen, and  $d/b=4$ ($\lambda$ is the wavelength).  The axes have arbitrary units.}
\label{fig:dst}
\end{figure}

The double-slit interference pattern is well described by the classical theory of light waves. However, to describe this pattern for photons, we must appeal to a quantum-mechanical formalism. 
In the quantum-mechanical description, if a photon is incident on two apertures (slits) separated by a distance $d$, we can assume for simplicity that the photon's probability amplitudes arising from each slit are the same. According to Feynman,\cite{Feynman} if it is indistinguishable from which slit the photon emerges, then the probability amplitude for reaching a screen at a position $\vec{r}$ will be the sum of the probability amplitudes from each slit.
Aside from the magnitude of its probability amplitude, each photon acquires a phase $\vec{k}\cdot\vec{r}$, where $\vec{k}$ is the propagation vector with magnitude $k=2\pi/\lambda$ and $\lambda$ is the wavelength of the photon. 
In Fig.~\ref{fig:dst}, we show a schematic of the double-slit interference setup. We assume the common situation: The distance to the screen is much larger than the slit separation, and thus the rays leaving the slits to a common point on the screen are nearly parallel. Then, the path difference is $d\sin\theta$, where $\theta$ is the angle relative to an axis centered on the slits and perpendicular to the plane in which they are contained. 
To proceed with this analysis, it is best to use the Euler notation. The total probability of finding a photon at a point $\vec{r}_0$ on the screen is the square of the sum of the amplitudes from the top and bottom slit
\begin{equation}
P=\left|ce^{ikr_0}\left( e^{-ik(d/2)\sin\theta}+e^{ik(d/2)\sin\theta}\right)\right|^2\propto\cos^2\alpha,
\label{eq:Pds}
\end{equation}
where $c$ is a normalization constant, and
\begin{equation}
\alpha=\frac{\pi d \sin\theta}{\lambda}\simeq \frac{\pi d x}{\lambda L}.
\label{eq:alpha}
\end{equation}
The screen is a distance $L$ away from the slits, and the point on the screen where we calculate the probability is a distance $x$ from the screen's center. Because $\theta\ll 1$, we have applied the small angle approximation: $\sin\theta\sim\theta\sim x/L$.

In the previous situation, the photon interferes with itself in going through both slits. However, if the difference in path to the screen from different points in the aperture of each slit is comparable to the wavelength, then we need to consider that the photon also interferes with itself in going through the various points of the aperture. 
Because each aperture is continuous, we must perform a continuous sum, i.e., an integral, in calculating the probability. 
\begin{equation}
P=\left|Ce^{ikr_0}\left( \int_{-d/2-b/2}^{-d/2+b/2} e^{iky\sin\theta}dy+\int_{d/2-b/2}^{d/2+b/2} e^{iky\sin\theta}dy\right)\right|^2\propto\left(\frac{\sin\beta}{\beta}\right)^2\cos^2\alpha,
\end{equation}
where $C$ is a normalization constant and 
\begin{equation}
\beta=\frac{\pi b \sin\theta}{\lambda}\simeq \frac{\pi b x}{\lambda L}
\label{eq:beta}
\end{equation}

When we have many photons, the total number of photons per unit time reaching the screen is
\begin{equation}
N=N_0P=N_m\left(\frac{\sin\beta}{\beta}\right)^2\cos^2\alpha=N_m\left(\frac{\sin\beta}{\beta}\right)^2\frac{1}{2}(1+\cos2\alpha),
\label{eq:IP}
\end{equation}
where $N_0$ is the number of photons per unit time incident on the apertures, and $N_m$ is the maximum number of photons per unit time on the screen. The form of the last relation is useful for accounting for experimental imperfections, as will be shown in Sec.~\ref{sec:cons}. The classical intensity is given by $ I=N_0EP$,
where $E=hc/\lambda$ is the energy of each photon, $h$ is Planck's constant, and $c$ is the speed of light in vacuum. This last relation agrees with the classical one.\cite{Bennett}  Because the problem reduces to two dimensions, the intensity has units of energy per unit time and length. The probability has units of inverse length.

A graph of the full pattern according to Eq.~(\ref{eq:IP}) is shown in Fig.~\ref{fig:dst}.
It can be thought of as consisting of a double-slit interference pattern [i.e., the term with  $\cos^2\alpha$ in Eq.~(\ref{eq:IP})] due to the two slits separated by $d$, with an amplitude or envelope determined by the single-slit diffraction pattern due to each slit of width $b$. Two-slit interference patterns are usually understood in terms of the maxima seen in the interference pattern, given by the classic formula $d\sin\theta=n\lambda$, where $n$ is an integer. The minima of the envelope are given by $b\sin\theta=m\lambda$, where $m$ is an integer. The first minima of the envelope (i.e., when $\beta=\pm\pi$, or equivalently, $x_{m=\pm 1}=\pm\lambda L/b$) delimits how many maxima of interference are seen in the central portion of the pattern. The number of maxima within this portion is roughly $2d/b-1$, as shown in Fig.~\ref{fig:dst} for the case $d/b=4$. 

If we only have a single slit, the photon still interferes with itself going through different parts of the single slit. The pattern exhibits the characteristic single-slit diffraction given by
\begin{equation}
N=N_m\left(\frac{\sin\beta}{\beta}\right)^2.
\label{eq:ss}
\end{equation}

In the previous double-slit situation, not being able to tell which slit the photon might go through makes the passage through the two slits indistinguishable,
and this produces a superposition of possibilities and, thus, interference. If the photons passing through the slit have information on which slit they went through, then the passage through the slits is distinguishable, and so there should be no superposition and, thus, no interference.\cite{Feynman} If we put orthogonally oriented polarizers over each of the slits, then the information of the slit passage will be encoded in the polarization, and interference will not occur. The probability of detecting a photon on the screen will be the sum of the probabilities from each slit considered separately.

For the sake of brevity, we will refer to the transmission axis of the polarizer as the \emph{orientation} of the polarizer. We have two polarizers, one in front of each slit, one oriented at $+45^\circ$ from the vertical and the other at $-45^\circ$ to the vertical. We now put a polarizer after the slits.  If the polarizer is oriented $+45^\circ$ or $-45^\circ$ from the vertical, there will not be interference because the light reaching the screen comes from only one slit---we need light from both slits to see interference. However, if the polarizer is oriented vertically, then a photon emerging from either slit will have $1/\sqrt{2}$ amplitude of going through the polarizer and, after the polarizer, the emerging photon has a polarization parallel to the transmission axis of the polarizer regardless of its previous polarization state. The polarizer projects the state of the photon. The slit-passage information is erased, and there is interference. This is the quantum eraser,
 where knowledge of the path information determines whether interference is present or not.\cite{ScullyN91}

The skeptical student may argue that the previous expression for double-slit interference is purely classical. Indeed, in this case, classical optics gives us correct answers when we have a lot or a few photons. Is there a slit interference situation that is not explained by classical optics? The answer is yes. When two or more identical photons are involved, new situations arise. If we send {\em both} photons together through the slits and record both photons, then the indistinguishable possibilities are doubled as follows: Both photons going through separate slits and both photons going through the same slit (two cases each). The interference pattern has an unusual shape,\cite{BrendelPRL91,FonsecaPRL99,DangeloPRL01} distinct from the single-photon pattern. The lab source used in the demonstration presented here produces photon pairs and can be configured that way.\cite{GalAJP05} Also, the photons of this source are entangled in momentum, and if set up appropriately, they can be entangled in the polarization degree of freedom. Because of this entanglement, a measurement of the entangled partner can modify the joint pattern seen,\cite{RibeiroPRA94} or even define whether interference is seen or not.\cite{WalbornPRA02,NevesNJP09} Taking the momentum correlation between two photons to the extreme, one can send one of the photons to the double-slit but image the pattern with the {\em other} photon, a type of imaging known as ``ghost'' imaging.\cite{StrekalovPRL95,AspdenAJP16}

\section{Apparatus}\label{sec:app}

Our apparatus has the core components of a standard correlated-photon layout.\cite{GalvezQuantumFest23} The basic setup consists of a pump diode laser of wavelength 405 nm incident on a beta-barium borate (BBO) crystal that enables spontaneous parametric down-conversion (SPDC). Photon pairs correlated in energy and momentum leave the crystal and these degenerate photons have a wavelength of 810 nm. After traveling through a table-top optical breadboard, the photons were collected into optical fibers that channeled them to single-photon detectors. Coincidence detection ensures that the recorded photons have single-photon statistics via heralding.\cite{MeyerRSI20} An attenuated laser is not a good source of single photons. Demonstrations with attenuated beams, even when using single-photon cameras,\cite{RuecknerAJP13} display discrete detections strongly suggesting single-photon events, but cannot be guaranteed to involve one-photon-at-a-time events.\cite{ThornAJP04} 

\begin{figure}[h!]
\centering
\includegraphics[width=5in]{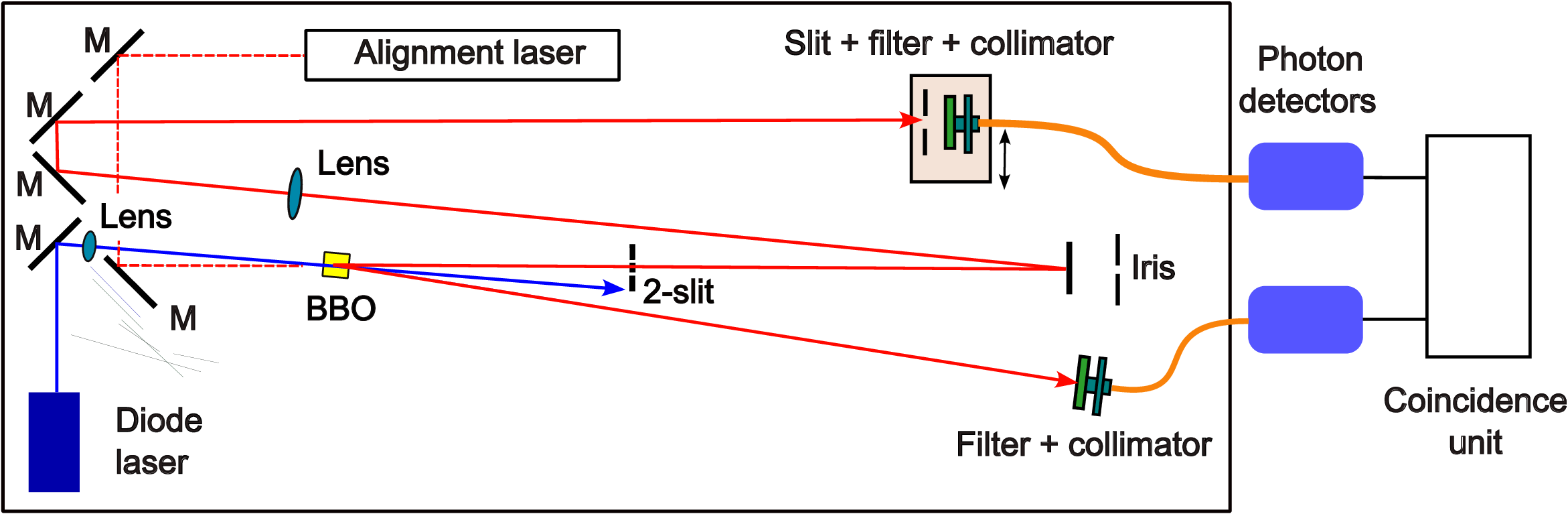}
\caption{(Color online) Schematic of the apparatus to demonstrate double-slit interference with single photons. A diode laser was focused on a down-conversion crystal (BBO) producing photon pairs. One went to a detector heralding the other photon that went toward a slit reticle. The photons emerging from the slit(s) were steered through mirrors (M). They traveled a total distance of 3 m in free space, passing through a lens of 1.5 m focal length halfway. A slit aperture delimiting the photons entering an optical fiber, and sending them to a detector, was scanned along a transverse plane via a motorized translation stage. The coincident detection of photon pairs was recorded.}
\label{fig:app}
\end{figure}

The double-slit additions to the setup are shown in Fig.~\ref{fig:app}. They include two lenses, one for focusing the pump laser onto the BBO crystal (best results with focal length $f_0\sim25$ cm), and another with long focal length (best results with $f\sim1.5$ m) between the slits and the photon collection; a double-slit reticle used for educational purposes (best results with $d\sim 0.6$ mm and $d/b\sim 4.7$); 3 translation stages: one to move the double slits along a transverse axis, and two to move the fiber collimator collecting the photons coming from the slits;  3 square mirrors to route the light over a 3-m ($2f$) path on the optical breadboard; and 3 (optional) infra-red polarizers for implementing the quantum eraser.  The values of some of the parameters used with these components were critical in the implementation. The criteria for their selection are discussed in detail in Sec.~\ref{sec:cons}.

\section{Experimental results}\label{sec:results}
We did three types of experiments---interference with double slits, single-slit diffraction, and the quantum eraser. We did numerous scans searching for the optimal parameters that would give the fastest and easiest demonstration.

\subsection{Double-slit Interference}
\begin{figure}[h!]
\centering
\includegraphics[width=5in]{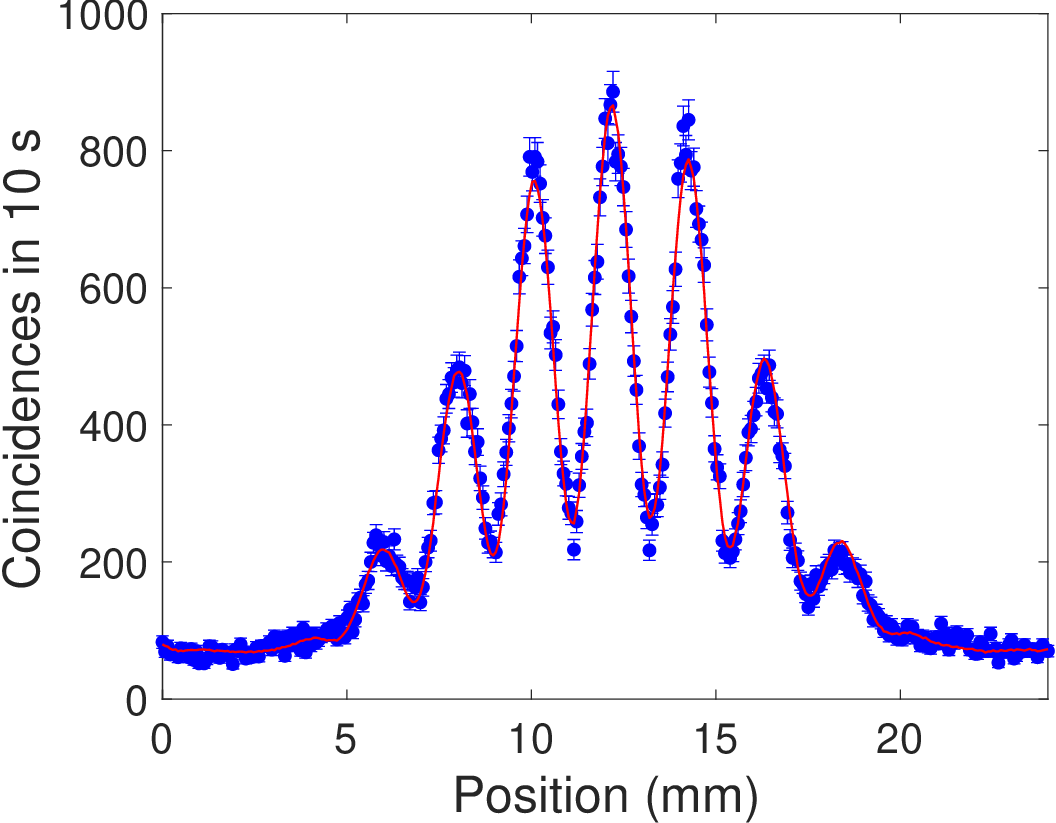}
\caption{(Color online) Double-slit interference pattern showing the coincident detections as the photon-collecting fiber arrangement was scanned transversely. The data shown were taken for $d=0.62$ mm and $b=0.13$ mm. The solid line is a fit of Eq.~(\ref{eq:Nv}) to the data.}
\label{fig:dslit}
\end{figure}

With a pump laser of about 30 mW and the setup of Fig.~\ref{fig:app} we acquired typical singles counts of about 3,000 per second for the photons that passed through the slits; 35,000 per second for photons that went straight to a detector; and about 7 coincidences per second. A minimum of 3 seconds per point was enough to get a recognizable interference pattern in about 15 minutes. Figure~\ref{fig:dslit} shows coincidence data set taken with a slit width of 0.7 mm on the scanning collection involving 341 data points over close to 25 mm. These data were taken for 10 s per point, taking about 21 minutes to complete. 
We make available a speed-up video of a similar scan as supplementary material.

The solid line in Fig.~\ref{fig:dslit} is a fit of Eq.~(\ref{eq:Nv}) to the data. The fit gives $d=0.60\pm0.02$ mm and $b=0.13\pm0.02$ mm, which are consistent with the measured values. Other fit parameters are given in Sec.~\ref{sec:cons}. 

To verify the single-photon aspect of the experiment, we added a beam splitter deflecting part of the light emerging from the slits to a detector and also recorded the triple coincidences. We used these data to calculate the degree of second-order coherence of the light, obtaining $g_2(0)=0.75\pm0.03$. A light source exhibiting quantum statistics should have $g_2(0)<1$, while classical-wave source requires $g_2(0)\ge 1$.\cite{ThornAJP04} This test gave us confidence that the experiment reflected a single-photon interference situation. We note that the measured value of $g_2(0)$ is high compared to typical values for single-photon experiments (typically less than 0.1).\cite{ThornAJP04} The slits produced such losses in the efficiency of detecting photon pairs that, in combination with long integration times, led to larger accidental coincidences relative to the real down-converted coincidences, giving the higher value of $g_2(0)$.

\subsection{Single-Slit Interference}
\begin{figure}[h!]
\centering
\includegraphics[width=5in]{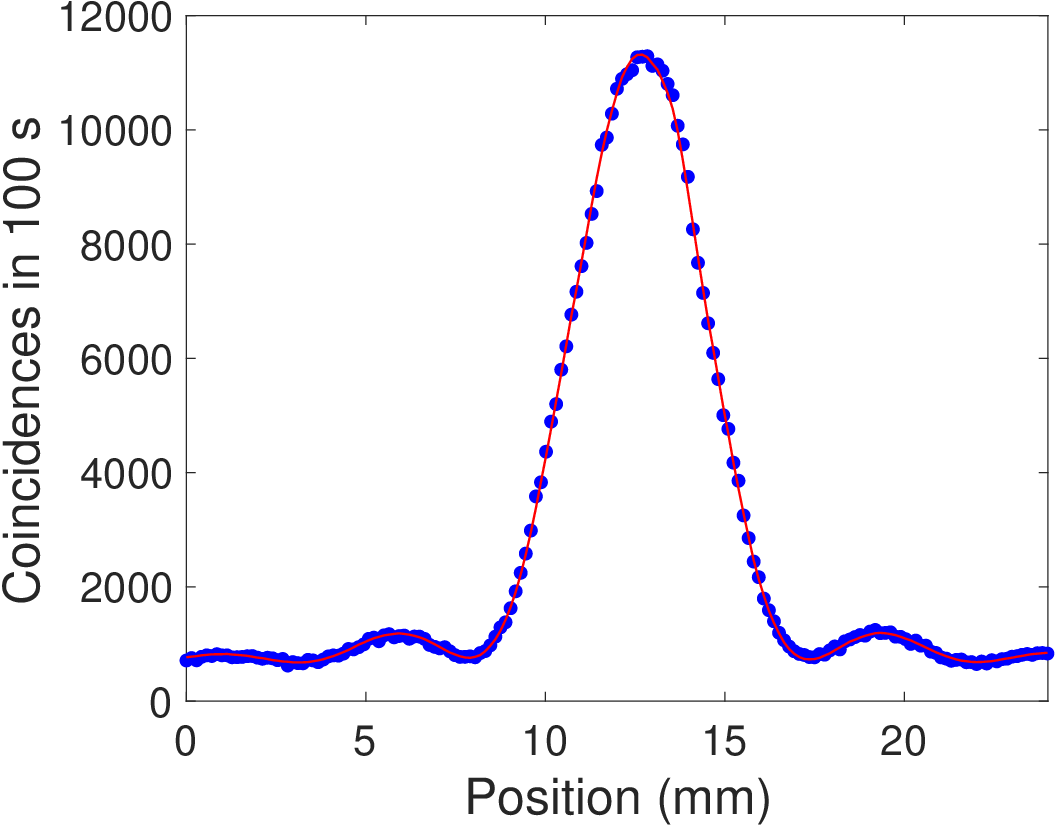}
\caption{(Color online) Measured interference pattern for a single slit with $b=0.285$ mm. The solid line is a fit of Eq.~(\ref{eq:ss}) to the data. Error bars were of the order of the size of the symbols and were omitted for the sake of clarity. }
\label{fig:sslit}
\end{figure}

We also took single-slit patterns. Figure \ref{fig:sslit} shows the data for one slit plus fit of Eq.~(\ref{eq:ss}).
 The fit gave $b=0.27\pm0.01$ mm, which is consistent with what we measured at $0.285\pm0.005$ mm.
The pattern is impressive because it agrees with the predicted maxima for first second and order, of 4.7\% and 1.6\% of the zero-order maximum, respectively. It demonstrates that photons are interfering with themselves in going through different portions of the single slit. 

\subsection{Quantum Eraser}
The third type of demonstration that can be performed with this platform is the quantum eraser, as described earlier. Unfortunately, it uses film polarizers that produce a lot of scattering, reducing the signal by a significant factor. 

\begin{figure}[h!]
\centering
\includegraphics[width=4in]{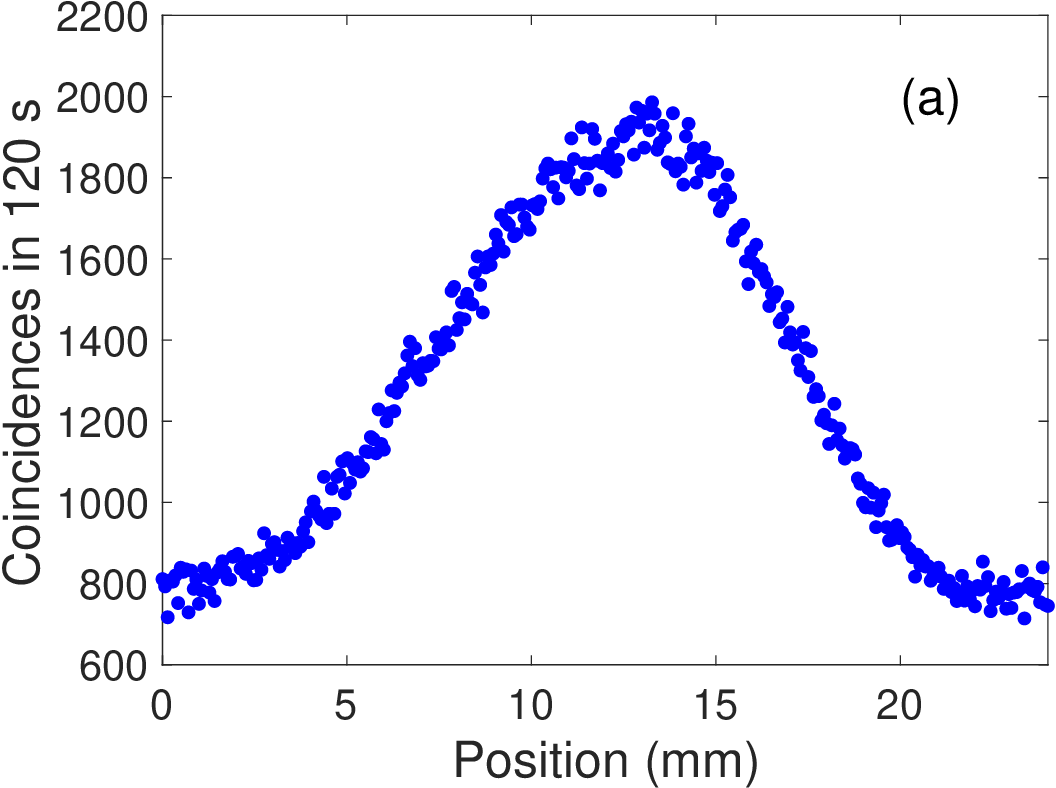}

\includegraphics[width=4in]{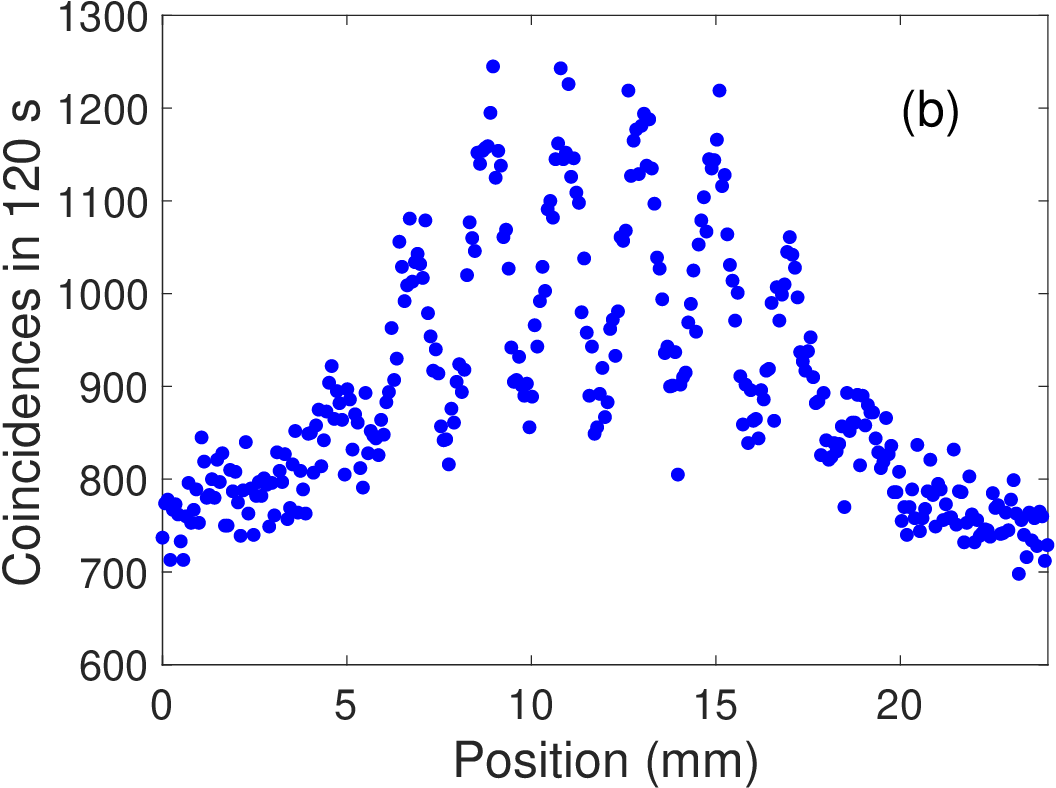}

\caption{(Color online) Graphs showing two sets of data for the quantum eraser situation: in (a) each slit was covered by a film polarizer, and with the axes of the polarizers being $+45^\circ$ to vertical for one slit and $-45^\circ$ to vertical for the other;  (b) was taken with a third polarizer placed after the slits with its transmission axis vertical.}
\label{fig:eraser}
\end{figure}

Figure~\ref{fig:eraser} shows two stages of the eraser. The input polarization was vertical. One slit had a polarizer at $+45^\circ$ to the vertical, and the other slit had a polarizer at $-45^\circ$ to the vertical. Thus, after the photons emerged from the slits, the photons carried path information (i.e., which slit they went through), and so there was no interference, as shown in Fig.~\ref{fig:eraser}(a). Notably, we did not need to measure the path information. Interference is not present whenever the path information is available, regardless of whether it is collected or not. When we put a polarizer after the slits oriented with the vertical direction,  photons from both slits had a probability amplitude of $1/\sqrt{2}$ of getting transmitted and, after doing so, their polarization was vertical carrying no memory of their previous polarization state and also containing no path information. As a consequence, interference appears, as shown in Fig.~\ref{fig:eraser}(b). If, 
instead, we orient the polarizer in the  
$+45^\circ$  or  $-45^\circ$  directions, we see no interference (not shown) because the data correspond to photons going through only one slit. 

\section{Implementation Considerations}\label{sec:cons}

\subsection{Pattern Scanning and Measurement}
The first consideration is the size of the ``screen'' where the pattern will be measured. It is not feasible to have a real screen as 
 is done in typical light-interference demonstrations 
 because we need to detect single photons. An electronic camera is an ideal solution, but for a table-top demonstration, current cameras are expensive (in the range \$40k--\$80k) and require a significant optical layout (an insertion delay of 20--40 ns requires the photons going to the camera to travel 20--40 ft in free space before reaching the camera).\cite{AspdenAJP16} Thus, we are left with scanning a single-photon detector. We do not scan the actual detector but the optical fiber that sends the photons to the single-photon detectors. For this operation, we needed a translation stage, and a reasonable span for such a device is 25 mm. We also wanted to automatically scan the position of the fiber, so we needed a 
 motorized screw that could push the stage over that distance. We stacked a second (manual) translation stage of a shorter span 
 on top of the motorized stage 
 to make fine adjustments to the center of the interference pattern. We also mounted the collimator on a vertical stage for fine (optional) adjustment of the vertical position.

\begin{figure}[h!]
\centering
\includegraphics[width=5in]{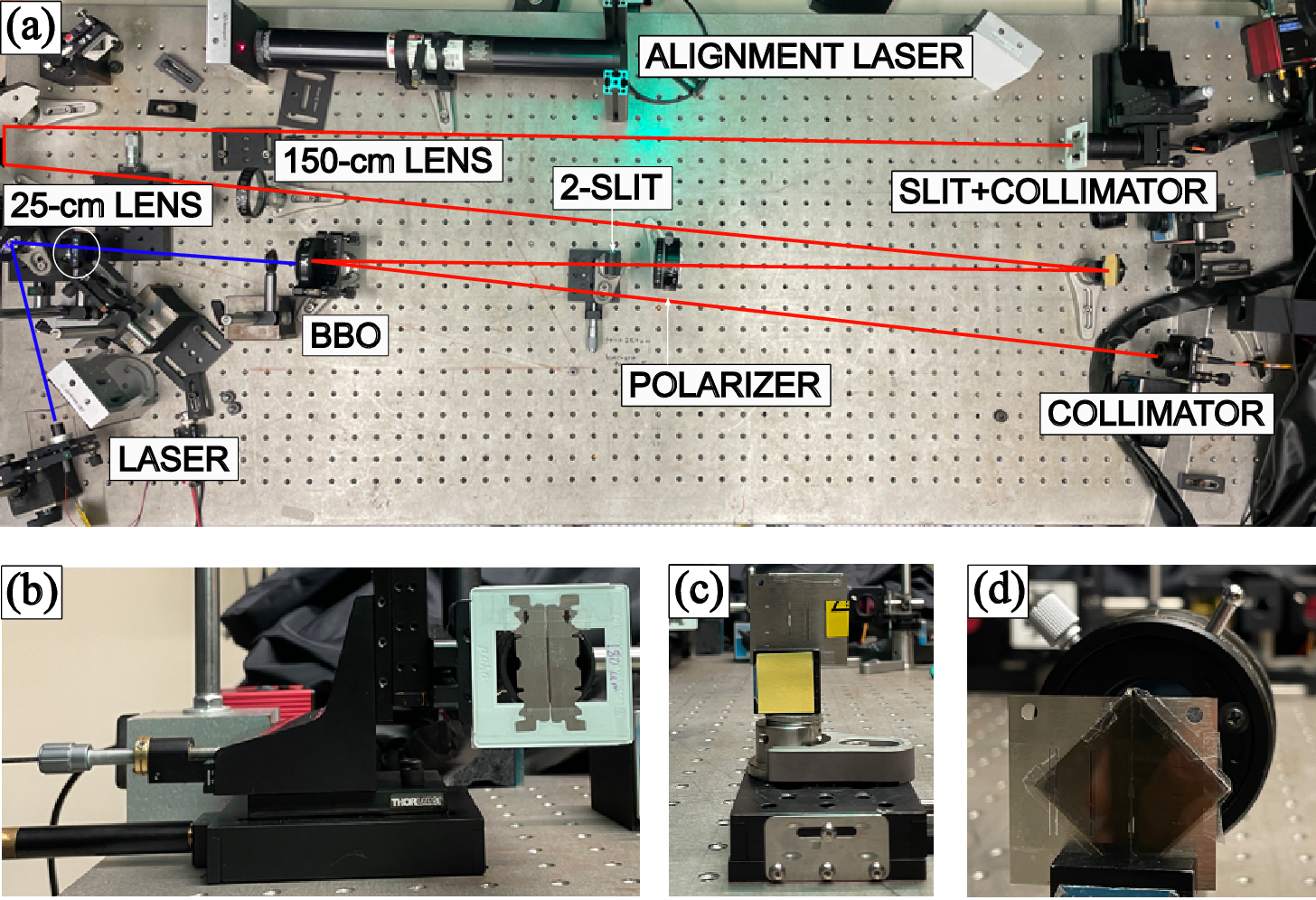}
\caption{(Color online) Photos of the entire setup (a) arranged over a 2-ft$\times$5-ft optical breadboard. Colored lines show the path of the light. The interference pattern was collected by a fiber and collimator mounted on two translation stages (one manual and one motorized), preceded by a mounted slit and filter (b). We used a clear aperture double-slit reticle (c) for the experiments. Two halves of a film polarizer blocked each of the slits for the quantum eraser experiment (d), with the eraser/polarizer mount seen in the background. }
\label{fig:photos}
\end{figure}

As seen in Fig.~\ref{fig:dst}, the most observable part of the pattern is within the first minima of the single-slit envelope. To fit this portion of the pattern within our detection span, using Eq.~(\ref{eq:beta}), we selected the distance between the two envelope minima given by 
\begin{equation}
\Delta x=x_{m=+1}-x_{m=-1}=2\lambda L/b,\label{eq:Dx}
\end{equation}
to be $\Delta x\sim 20$ mm. A 
compelling interference pattern should include a few interference fringes. If we pick $d/b\sim4$, the fourth interference maximum is suppressed by the minimum of the envelope, so we would get seven visible fringes within the envelope, as shown in Fig.~\ref{fig:dst}. The fringe separation is then
\begin{equation}
\delta x=x_{n+1}-x_{n}=\lambda L/d,
\end{equation}
which in combination with Eq.~(\ref{eq:Dx}), gives $\delta x= \Delta x \;b/(2d)\sim 20/8=2.5$ mm. One factor affecting the resolution of the scan is the aperture of the photon-collection optics. 
To observe fringes with good visibility,  
our aperture should be 
$a\ll \delta x$. Initially, we made a single-slit opening with old-fashioned razor blades with a separation of $a\sim0.18$ mm, as shown in Fig.~\ref{fig:photos}(b), with which we took a sizable amount of data. Later, using a slit of variable width, we found that we could use a maximum width opening of 0.7 mm without affecting the resolution, which was dominated by the spatial coherence discussed in Sec.~\ref{sec:coh}.

\subsection{Pattern Expansion and Focusing}
To measure the main pattern over a 25-mm span we needed about 1.5-m distance between the slits and the ``screen.'' To ensure that the light would be incident on the collimator at near normal incidence, the ``screen'' was located a focal length $f$ away from the lens. This doubled the slit to ``screen'' distance to $2L=2f\sim 3$ m. We used a lens with $f\sim 1520$ mm and a diameter of 51 mm, which is almost the maximum span we could reasonably arrange for the beam to travel on the breadboard. With this arrangement, the detected pattern is the Fourier transform of the slits. (We also tried a 4-m span with $f\sim 2$ m, but we did not get the advantage that we expected for the trouble, so we abandoned it.)  As seen in Fig.~\ref{fig:app}, we used three extra mirrors (M) to steer the beam through the 3-m distance from the slit to the ``screen.'' The first folding mirror was a ``flip'' mirror, which was folded out of the way for the initial alignment to get down-converted photons. An iris (I$_{\rm R}$) was used to mark the original trajectory of the down-converted photons. We had an alignment laser beam set up to follow the path of the down-converted photons (i.e. traveling through the same location where the pump beam traversed the BBO crystal and iris I$_{\rm R}$). This alignment laser was also critical for setting the optics that steered the down-converted photons to the detectors.

\subsection{Slit Reticle}
From our calculation, the width of the slits should be $b=2\lambda L/\Delta x=0.12$ mm, and their separation $d=4b=0.48$~mm. We tried several double-slit reticles sold for educational purposes. Very few had dimensions that were close to those needed. The best one (Leybold model 46992) is made of stainless steel with clear apertures, with $b\simeq 0.13$ mm and $ d\simeq 0.62$ mm, or $d/b=4.8$, which gave seven fringes within the envelope. Backtracking through the calculations, this gave us an envelope stretching $\Delta x=19$~mm, with a fringe separation $\delta x=2$~mm.

The slit reticle was mounted on a translation stage. We found out that transverse adjustment was needed to center the slits on the path of the down-converted beam. Slight misalignments would produce uneven amplitudes between the two slits, decreasing the visibility of the fringes.  Figure~\ref{fig:photos}(c) shows a photo of the slit on its translation-stage mount. Not shown is an arrangement of a mask to block only one slit. We used this in conjunction with the stage to make sure we got about the same number of coincidences from each slit. We also used a less complicated method that involved setting the detector in a fixed position in the middle of its scanning range and manually scanning the double slit. A central maximum plus two adjacent secondary maxima of interference were recorded. Setting the double slit at the peak of the central maximum ensured it was best aligned. 

We also used one of the slits in the same reticle to observe a single-sit pattern. Because we wanted to see the secondary maxima within our measurement range, we needed a larger opening than the one used. One of the pair of slits of wide separation had slits of width $b=0.25$ mm, so we blocked one of the slits and used the other one as a single-slit aperture. This produced a pattern where the separation between the second diffraction minima was predicted to be about 10 mm, as seen in Fig.~\ref{fig:sslit}.

\subsection{Spatial Coherence and Resolution}\label{sec:coh}
The size of the source of the photons incident on the slits is the most relevant factor because it relates to the spatial coherence of the input light. In simple terms, different points from the source produce patterns overlapping on the screen to blur the whole pattern. The spatial coherence effect was originally realized by Michelson in 1890,\cite{Michelson1890} which he used to measure the diameter of Betelgeuse.\cite{MichelsonPNAS21} 
In our case, the down-conversion source is spatially incoherent, so the size of the pump beam on the BBO crystal will also contribute to the visibility of the interference pattern or sharpness of the fringes.

The slits were located at $z=30$ cm from the BBO crystal. The angular size of the source should not be larger than the desired angular resolution. For example, if we set  $a/L\sim1.6\times 10^{-4}$  by picking $a=0.25$~mm, our resolution would be 1/10 of the fringe separation. We measured the half-width of the pump laser beam to be $\omega=0.52$ mm. If we just send the laser beam of size $A\sim2\omega= 1$ mm to the BBO crystal, the angular spread of the source of down-converted photons would be $A/z=3.3\times 10^{-3}$, which is about 20 times larger, or 2 times the fringe separation, not making the pattern visible at all. To reduce the size of the beam, we must focus the beam. If we focus the pump beam with a lens with a focal length $f_0$, the convergence angle of a Gaussian beam is $\gamma=\lambda_0/(\pi \omega_0)$,\cite{Bennett} where $\lambda_0=405$~nm and $\omega_0$ is the focused waist. Using $\gamma=\omega/f_0$, 
and with $f_0 = 25$~cm  we get $\omega_0=0.064$~mm. This gives 
an angular size $4.3\times 10^{-4}$ that is of the order of our desired collecting resolution. By setting $a=0.7$~mm, mentioned earlier, it gave $a/z\sim 4.6\times 10^{-4}$, matching the angular size of the source. For higher values of $a$, the slit dominated the resolution, whereas for smaller values, the spatial coherence was the dominant factor.

To account for the coherence in the measured interference pattern, we must express the interference pattern by\cite{BornWolf}
 \begin{equation}
 N=N_0\left(\frac{\sin\beta}{\beta}\right)^2\frac{1}{2}\left[1+|V|\cos(2\alpha+\delta)\right]
 \label{eq:Nv}
 \end{equation}
 where $V$ is the visibility of the pattern or the complex degree of spatial coherence, and $\delta$ is a phase. $V$ is calculated by the van Citter-Zernike theorem, which states that the visibility of the pattern is the Fourier transform of the aperture function.\cite{BornWolf} In our case, the source is formed by the waist $w_0$ of the pump beam at the BBO crystal, which is a Gaussian. The visibility is then\cite{RibeiroAO94}
 \begin{equation} 
 |V|=\exp\left[-\frac{(\pi d w_0)^2}{(\lambda z)^2}\right].\label{eq:v}
 \end{equation}
Using the parameters of the apparatus, we calculate $V=0.77$. 
In the fit of Eq.~(\ref{eq:Nv}) to the data in Fig.~\ref{fig:dslit}, the visibility comes out to be $|V|=0.65\pm0.22$, which is slightly lower than the estimate based on Eq.~(\ref{eq:v}). Using a shorter focal length does not help because, although the waist of the beam is much smaller, its width along the length of the BBO crystal is not constant due to having a Rayleigh range comparable to or smaller than the crystal length. Thus, the visibility obtained with the 25-cm lens is good enough to serve as a demonstration. 

Once the light was focused, we found that the resolution-limiting effect was the projection of the excited region of the BBO in the crystal, which is not accounted for in the preceding analysis. That is, the source of photons is not a point but a line. Using a 3-mm crystal and detecting photons at 3$^\circ$ to the pump direction results in a maximum effective source size of about 3-mm$\times \sin 3^\circ=160\;\mu$m in size. We have verified that a thinner 0.5-mm crystal and a smaller down-conversion angle alleviate this problem, obtaining $|V|=0.76\pm0.20$. However, the thinner crystal implies a smaller signal and longer integration times, making it a time-consuming demonstration. All the data presented here were obtained using the 3-mm crystal and 3$^\circ$ angle. We did not account for the spatial coherence as it applied to the slit widths, although it is expected to be a higher-order effect.\cite{PearsonOSAC18}

\subsection{Down-converted Bandwidth}
Since our resolution is about 35\% of the fringe separation, the bandwidth should play a role only when it gets above 35\% of the wavelength, which is well above the value of the filters normally used in this setup (30 nm, centered at 810 nm). To check this, we tried 10-nm filters, and they did not improve the resolution and only reduced the overall photon counts.

\section{Conclusions and Discussion}\label{sec:conc}
In summary, we discussed a setup to demonstrate Young's double-slit experiment with single photons. We present a laboratory arrangement and components that are feasible to implement using a table-top correlated photon apparatus. We investigated the range of parameters that allow easy implementation without major modifications of the down-conversion setup. 
Critical 
issues
 include the need to focus the pump beam and the expansion of the beam after the slits. 
We also presented a version of the quantum eraser with polarization. The outcomes go to the heart of quantum interference and complementarity. 

This type of slit interference has been found in systems other than photons, such as electrons,\cite{BachNJP13} atoms,\cite{CarnalPRL91} molecules,\cite{BrandAJP21} and neutrons,\cite{ZeilingerRMP88} yet the physics is the same for all. We should then invoke the {\em quanton} as the quantum object that experiences the quantum interference situation regardless of the actual physical carrier. 
The search for a quantitative representation of complementarity introduced distinguishability and visibility as measurable complementary quantities.\cite{WootersPRD79} This approach has gained much recent interest, by establishing an inequality linking the amplitudes of the photon going through each slit with a {\em which-way} marker for each slit; a form of entanglement between the light and the apparatus.\cite{EnglertPRL96} 
This marker can be, for example, the polarization of the light.\cite{EnglertZN99} Extensions of this understanding of the interference include the degree of coherence of the light.\cite{EberlyOptica17,DeZelaOptica18} 
Quantification of complementarity is manifested in the visibility of the interference fringes, which depends on the which-way marker and the degree to which the paths are distinguishable or incoherent. 
The complementary behavior goes beyond the wave and particle situation. It is present on any pair of incompatible observables, such as spin components along orthogonal axes. It is an intrinsic property of quantum systems. 

 A more recent development on slit interference involves ``looped paths'' in slit interference. That is, the quanton may propagate out of one slit, back in through a neighboring slit, and back out of a different slit, such as in the case of three slits, or two slits with only one slit illuminated.\cite{SawantPRL14}  Those paths are enhanced by interactions of the quanton and slits via plasmons in the case of light.\cite{MaganaLoaizaNC16} Such situations may provide stringent tests to the Born rule of quantum mechanics.\cite{SinhaSci10}

\appendix

\section{Parts List}

Table~\ref{tab:basicparts} has the core components of the apparatus for generating and detecting photon pairs. It does not include accessories (irises, mirrors) nor mounting hardware (posts, mounts, clamps). More information on these components is given in Ref.~\onlinecite{URL}. Table~\ref{tab:parts} lists the main parts that are needed to expand the correlated-photon setup for this demonstration. Other parts can be obtained from Ref.~\onlinecite{URL}. 

 \begin{table}[p]
\centering
\caption{Parts list of main components of the basic setup with specific vendors, models, and prices. The list does not include mounting hardware for the components (see Ref.\onlinecite{URL} for details). There are alternative vendors for all the components.}
\begin{ruledtabular}
\begin{tabular}{c p{2.5cm} c p{4.5cm} r p{5cm}}
Item & Name & Number & Vendor \& Model & Price (\$) & Comment \\
\hline	
1 &Laser & 1 & Power Technology GPD(405-50)& 600  & 405 nm GaN laser.\\
2 & Crystal & 1 & Newlight Photonics NCBBO05300-405(I)-HA3& 550 & BBO type-I, 3 mm.\\
3 & Detectors & 2 & Excelitas/Alpha SPCM-EDU& 1700 &Single photon detector modules.\\
4 & Coincidence electronics & 1 & Altera DE2-115 or Red Dog & 350 & FPGA-based processing electronics.\\%
5 & Optical Fibers & 2 & Thorlabs M31L01&50 & Multimode optical fiber.\\
6 & Collimator & 2 & Thorlabs F220FCB & 150 & For focusing light into fibers.\\
7 & Filters & 2 & Newlight Photonics NBF810-30 & 160 & 30-nm bandpass filters for 810-nm.\\
8 & Optical Breadboard & 1 & Thorlabs B3060G& 1900 & 2-ft$\times$5-ft$\times$4-in.\\
\end{tabular}
\end{ruledtabular}
\label{tab:basicparts}
\end{table}

  \begin{table}[p]
\centering
\caption{Parts list of main components with specific vendors, models, and prices. Compatible components from other vendors or components available in-house work as well. Items 7, 8 and 9 are only needed for the quantum eraser demonstration.}
\begin{ruledtabular}
\begin{tabular}{c p{2.5cm} c p{4.5cm} r p{5cm}}
Item & Name & Number & Vendor \& Model & Price (\$) & Comment \\
\hline	
1 &Lens & 1 & OptoSigma SLB-50-1500PIR2 & 90  & 1.5-m, 2-inch diameter converging lens.\\
2 & Lens & 1 & Thorlabs model 1461& 65 & 25-mm, 1-inch diameter converging lens.\\
3 & Flip mount & 1 & Edmund Optics 19-409 & 95 &Mount to swivel a mirror.\\
4 &Mirror & 1 & Edmund Optics 68-380 & 115 & 25 mm square Silver mirror.\\%
5 &Mirror & 2 & Edmund Optics 68-334&190 & 75 mm square Silver mirror.\\
6 & Mirror mount & 3 & Edmund Optics 58-866 & 95 & For mounting mirrors.\\
7 & Polarizer & 2 & Edmund Optics 25-106 & 105 & Near-IR sheet. \\
8 & Polarizer & 1 & Edmund Optics 48-893 & 345 & Plate polarizer, near IR.\\
6 & Reticle slits & 2 & Leybold models 46992 and 46993 & 65 & Single, double \& multiple clear apertures on stainless steel.\\
9 & Rotational mount & 1 & OptoSigma GTPC-SPH30 & 250 & Manual rotation stage. \\
10 & Stage motor controller & 1 & Thorlabs KDC101 & 758 & Controller plus software.\\
11 & Translation stage & 2 & Thorlabs MT1 & 325 & Manual, with micrometer.\\
12 & Translation stage & 1 & Thorlabs PT1& 493 & 25-mm travel.\\

\end{tabular}
\end{ruledtabular}
\label{tab:parts}
\end{table}

\begin{acknowledgments}
We thank T. Campbell,  D. Chartrand, J. Freericks, and J. K\"uchenmeister for suggestions and encouragement.
This work was funded by National Science Foundation grant PHY-2011937.

\end{acknowledgments}

\end{document}